\documentclass[]{gHPR2e}
\usepackage{graphicx}
\usepackage{amssymb}
\usepackage{amsmath}

\advance\textheight by 2\baselineskip

\begin{document}
\doi{10.1080/0895795YYxxxxxxxx}
\issn{1477-2299}
\issnp{0895-7959} \jvol{31} \jnum{4} \jyear{20xx} \jmonth{xxxx}

\newcommand{\bk}{{\bf k}}
\newcommand{\bq}{{\bf q}}
\newcommand{\bQ}{{\bf Q}}
\newcommand{\bG}{{\bf G}}
\newcommand{\bK}{{\bf K}}
\newcommand{\bp}{{\bf p}}
\newcommand{\bx}{{\bf x}}
\newcommand{\by}{{\bf y}}
\newcommand{\br}{{\bf r}}
\newcommand{\bR}{{\bf R}}
\newcommand{\bJ}{{\bf J}}
\newcommand{\bz}{{\bf 0}}
\newcommand{\ba}{{\bf a}}
\newcommand{\bh}{{\bf h}}
\newcommand{\bd}{{\bf d}}
\newcommand{\bv}{{\bf v}}
\newcommand{\bdelta}{{\boldsymbol\delta}}
\newcommand{\bsigma}{{\boldsymbol\sigma}}
\newcommand{\Li}{{\mathop{\rm{Li}}\nolimits}}
\newcommand{\cotg}{{\mathop{\rm{cotg}}\nolimits}}
\renewcommand{\Im}{{\mathop{\rm{Im}}\nolimits\,}}
\renewcommand{\Re}{{\mathop{\rm{Re}}\nolimits\,}}
\newcommand{\sgn}{{\mathop{\rm{sgn}}\nolimits\,}}
\newcommand{\Tr}{{\mathop{\rm{Tr}}\nolimits\,}}
\newcommand{\EF}{E_{\mathrm{F}}}
\newcommand{\kB}{k_{\mathrm{B}}}
\newcommand{\kF}{k_{\mathrm{F}}}
\newcommand{\nF}{n_{\mathrm{F}}}
\newcommand{\vF}{v_{\mathrm{F}}}
\newcommand{\bvF}{\bv_{\mathrm{F}}}
\newcommand{\Green}{{G}}
\newcommand{\dSC}{{\mathrm{dSC}}}
\newcommand{\dPG}{{\mathrm{dPG}}}
\newcommand{\dDW}{{\mathrm{dDW}}}
\newcommand{\Ret}{{\mathrm{R}}}
\newcommand{\Tau}{T_\tau}
\newcommand{\modified}[1]{{\relax #1}}

\markboth{Pellegrino, Angilella, Pucci}{Transport properties across nonuniform
strain barriers in graphene}


\title{Ballistic transport properties across nonuniform strain barriers in
graphene}

\author{F. M. D. Pellegrino$^{\rm a,b}$,
G. G. N. Angilella$^{\rm a,b,c,d}$$^{\ast}$\thanks{$^\ast$Corresponding author.
Email: giuseppe.angilella@ct.infn.it\vspace{6pt}}
and R. Pucci$^{\rm a,b}$\\\vspace{6pt}  
$^{\rm a}${\em{Dipartimento di Fisica e Astronomia, Universit\`a di Catania, Via
S. Sofia, 64, I-95123 Catania, Italy}}; 
$^{\rm b}${\em{CNISM, UdR Catania, I-95123 Catania, Italy}}; 
$^{\rm c}${\em{Scuola Superiore di Catania, Via Valdisavoia, 9, I-95123 Catania,
Italy}}; 
$^{\rm d}${\em{INFN, Sez. Catania, I-95123 Catania, Italy}}
\\\vspace{6pt}\received{\today}}

\maketitle

\begin{abstract}

We study the effect of uniaxial strain on the transmission and the conductivity
across a strain-induced barrier in graphene. At variance with
conventional studies, which consider sharp barriers, we consider a more
realistic, \emph{smooth} barrier, characterized by a nonuniform, continuous
strain profile. Our results are instrumental towards a better understanding of
the transport properties in corrugated graphene.
\bigskip
\begin{keywords}
graphene; uniaxial strain; conductivity; 
\end{keywords}
\end{abstract}

Graphene is a single layer of $sp^2$ carbon atoms, arranged as an honeycomb
lattice. Its fabrication in the laboratory \cite{Novoselov:05a} immediately
stimulated the interest of both the experimental and theoretical communities,
and many applications, which initially could only be speculated, now appear
feasibile. In particular, electronic quasiparticles in graphene are
characterized by a band structure consisting of two bands, touching at the Fermi
level in a linear, cone-like fashion at the so-called Dirac points $\pm\bK$, and
a linearly vanishing density of states (DOS) at the Fermi level
\cite{CastroNeto:08,Abergel:10}. This implies in this novel condensed matter
system the possibility of Klein tunneling across barriers 
\cite{Katsnelson:06,MiltonPereira:06,Barbier:09,Barbier:10,Barbier:10a,Peres:09},
\emph{i.e.} perfect transmission across energy barriers, which was predicted in
the context of quantum electrodynamics at relatively much larger energies.

Graphene, like most carbon compounds, is also characterized by quite remarkable
mechanical properties. Despite its reduced dimensionality, graphene possesses a
relatively large tensile strength and stiffness \cite{Booth:08}, with graphene
sheets being capable to sustain elastic deformations as large as $\approx 20$\%
\cite{Kim:09,Liu:07,Cadelano:09,Choi:10,Jiang:10}. Larger strains would then
induce a semimetal-to-semiconductor transition, with the opening of an energy
gap
\cite{Gui:08,Pereira:08a,Ribeiro:09,Cocco:10,Pellegrino:09b,Pellegrino:09c,Pellegrino:11}.

Recently, it has been suggested that graphene-based electronic devices might be
designed by suitably tailoring the electronic structure of a graphene sheet
under applied strain (the so-called `origami' nanoelectronics)
\cite{Pereira:09}. Indeed, a considerable amount of work has been devoted to the
study of the transport properties in graphene across strain-induced single and
multiple barriers \cite{Gattenloehner:10}. It has also been suggested
that strain may induce relatively high pseudo-magnetic fields \cite{Guinea:10},
whose effects have actually been confirmed experimentally in graphene
nanobubbles grown on top of a platinum surface \cite{Levy:10}. Indeed, the
effect of the strain-induced displacement of the Dirac points in reciprocal
space can be (formally) described in terms of the coupling to a gauge field.
However, applied strain also induces a variation of the Fermi velocity, $\vF$.
In particular, uniaxial strain implies a Fermi velocity anisotropy, while an
inhomogeneous strain implies a nonuniform (\emph{i.e.} coordinate dependent)
velocity profile. This can also be realized in the presence of smooth potential
barriers, where it has been demonstrated that a nonuniform space variation of
the underlying gate potential would result in a modulation of the Fermi velocity
\cite{Cayssol:09,Concha:10,Raoux:10}. Moreover, there is considerable evidence,
both experimental \cite{Lee:08a} and theoretical \cite{Cayssol:09}, that barrier
edge effects are also important to determine the transport properties across
corrugated graphene.

Close to the Fermi energy and in the unstrained case, the electrons dynamics is
governed by the linearized Hamiltonian
\begin{equation}
H = \hbar\vF \bsigma\cdot\bp ,
\label{eq:H}
\end{equation}
where $\bsigma=(\sigma_x , \sigma_y )$ is a vector of Pauli matrices, associated
with the in-plane spinorial nature of the quasiparticles in graphene.
Eq.~(\ref{eq:H}) can also take into account intervalley processes $\bK
\leftrightarrow -\bK$, which however can be safely neglected, at sufficiently
low energies.

Applied strain is then described by means of the strain tensor
\cite{Pereira:08a} ${\boldsymbol\varepsilon} =  \frac{1}{2} \varepsilon
[(1-\nu){\mathbb I} + (1+\nu) A(\theta)]$, where $A (\theta)  = \sigma_z
e^{2i\theta\sigma_y}$. Here, $\varepsilon$ is the strain modulus, $\theta$ the
angle along which strain is applied, and $\nu=0.14$ is the Poisson ratio for
graphene \cite{Pellegrino:09b,Pellegrino:10a,Pellegrino:11,Pellegrino:11c}.
Starting from a more general, tight-binding Hamiltonian \cite{CastroNeto:08},
and expanding to first order in the strain modulus, one obtains an anisotropic
dependence on the strain angle $\theta$, already at linear order in the impulses
\cite{Pellegrino:11c}. This can mapped back to a linear Hamiltonian as in
Eq.~(\ref{eq:H}) \cite{Pellegrino:11}, where now impulses are reckoned from the
shifted Dirac points, and the Fermi velocity is anisotropic and possibly
coordinate-dependent \cite{Pellegrino:11c}.

We therefore consider a smooth strain barrier, characterized by a nonuniform,
continuous strain profile $\varepsilon = \varepsilon(\xi)$, with
\begin{equation}
\varepsilon(\xi) = \frac{\varepsilon_0}{\tanh(D/4a)} \left( \frac{1}{1+e^{-\xi/a}} -
\frac{1}{1+e^{-(\xi-D)/a}} \right) ,
\label{eq:smooth}
\end{equation}
where $\xi$ is the coordinate along the strain direction, forming an
angle $\theta$ with the crystallographic $x$ axis.
Such a strain profile is essentially flat for $|\xi-D/2|\ll a$, where
$\varepsilon(\xi)\approx\varepsilon_0$, and for $|\xi-D/2|\gg a$, where
$\varepsilon(\xi)\approx 0$. In the limit $a/D\to0$, Eq.~(\ref{eq:smooth}) tends
to a sharp barrier. 
The linear extent $a$, over which the strain profile Eq.~(\ref{eq:smooth})
varies appreciably, is naturally to be compared with the lattice step
${\mathfrak a}$, at the microscopic level, and with the Fermi wavelength
$\lambda_{\mathrm{F}} = \hbar v_{\mathrm{F}} /(2\pi E)$, where $E$ is the energy
of the incoming electron. While a smooth profile can be expected on quite
general grounds, the approximation of a sharp barrier is expected to hold well
whenever $a\ll {\mathfrak a}\ll \lambda_{\mathrm{F}}$, \emph{i.e.} at
sufficiently large incident energies. On the other hand, the details of the
strain profile come into their own when $a\sim\lambda_{\mathrm{F}}$.

Single electron tunneling, and thus the majority of the transport
properties of interest, can then be inferred by solving the stationary Dirac
equation associated to Eq.~(\ref{eq:H}), now including the nonuniform strain,
Eq.~(\ref{eq:smooth}) \cite{Pellegrino:11c}. For $|\xi|\to\infty$, the solutions
for the scattering problem are therefore known analytically
\cite{Pellegrino:11c} (and refs. therein). Integrating the scattering equations
from large positive $\xi$ backwards to large negative $\xi$, and comparing with the
known analytical solution, one may extract the reflection coefficient $r$,
relative to an incident wave with unit amplitude incoming from $\xi>0$, as the
Fourier weight with respect to its negative frequency component, whence the
transmission $T(E,\phi)$ at given incidence energy $E$ follows
straightforwardly.

\begin{figure}[t]
\begin{center}
\begin{minipage}[c]{0.3\columnwidth}
\centering
\includegraphics[width=\textwidth]{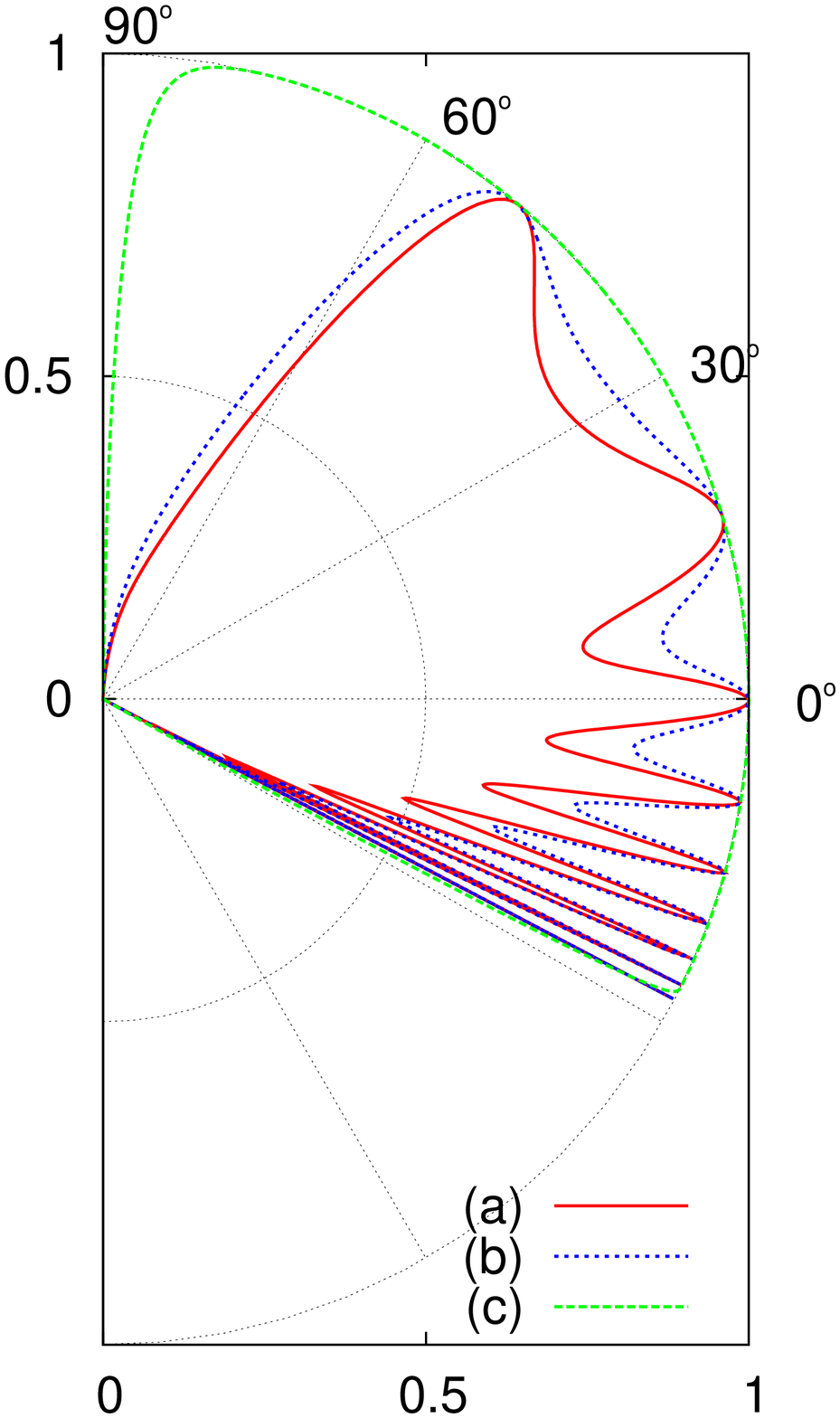}
\end{minipage}
\begin{minipage}[c]{0.6\columnwidth}
\centering
\includegraphics[height=\textwidth,angle=-90]{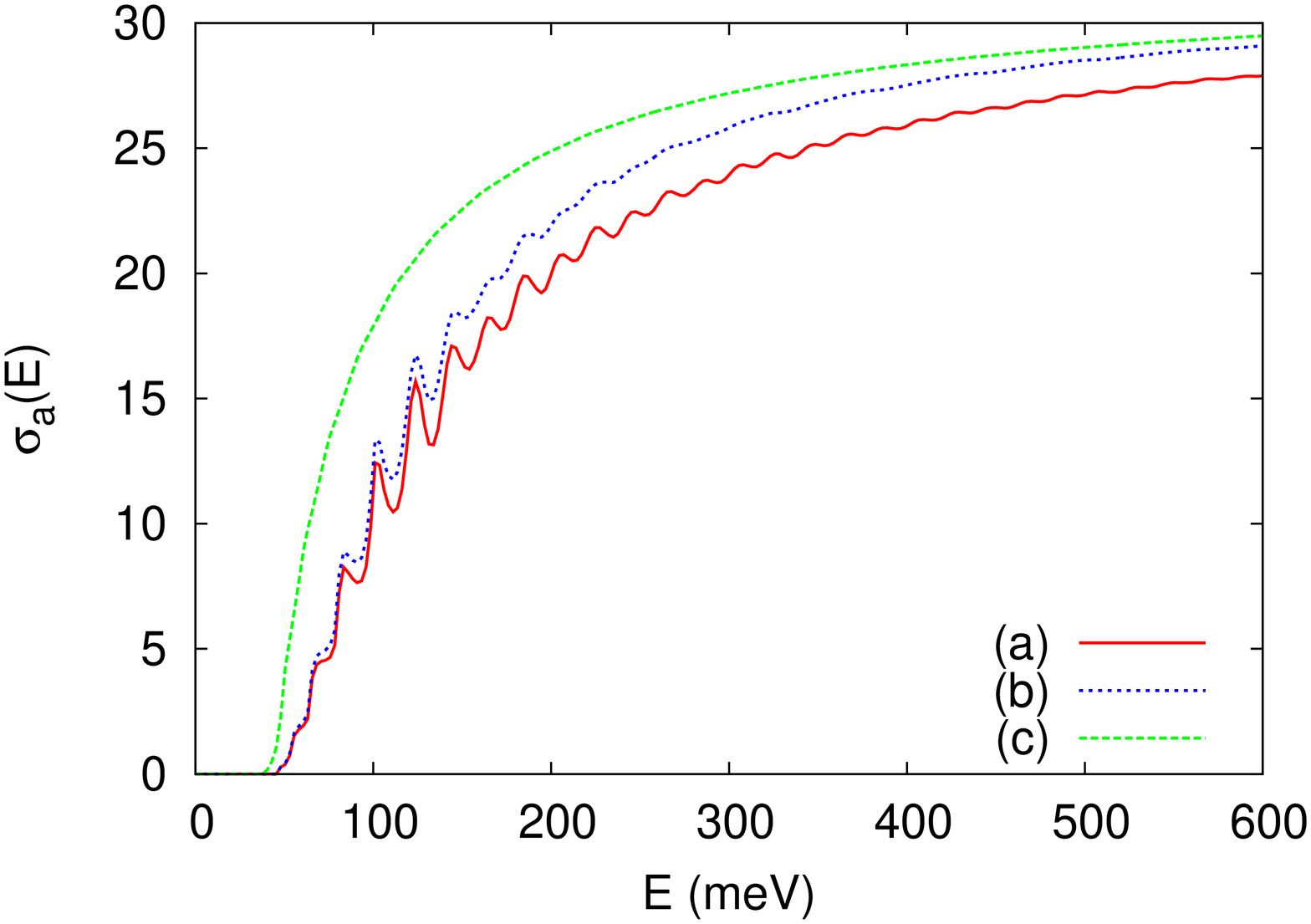}
\end{minipage}
\end{center}
\caption{(Color online) \emph{Left:} Tunneling transmission \emph{vs} incidence
angle $\varphi$ across a smooth strain barrier, with $D = 100$~nm, and incidence
energy $E = 167$~meV ($\lambda_{\mathrm{F}} = 0.6$~nm). \emph{Right:} Normalized
conductivity, $\sigma_a (E)$, \emph{vs} incidence energy $E$ across a smooth
strain barrier, Eq.~(\ref{eq:sigmaa}). In both panels, $\varepsilon_0=0.01$,
$\theta=\pi/2$ (strain is applied in the armchair direction), while different
curves refer to (a) sharp barrier, $a=0$; (b) $a=10^{-2} D$; (c)
$a=10^{-1} D$.}
\label{fig:condsmooth}
\end{figure}

Fig.~\ref{fig:condsmooth} (left panel) shows the transmission $T(E,\varphi)$ as
a function of the incidence angle $\varphi$ across strain-induced sharp and
smooth barriers, Eq.~(\ref{eq:smooth}), with strain applied along the armchair
direction ($\theta=\pi/2$). One observes that, upon increasing the smoothing
parameter $a/D$, the oscillations, characteristic of Klein tunneling across
energy barriers in graphene, get damped, while their envelope (lower bound)
increases.
The dependence of the transmission $T(E,\varphi)$ on the incidence angle
$\varphi$ is only apparently asymmetric, as we are restricting to quasiparticles
with momentum centred around a given Dirac cone, say $+\bK$. Symmetry
$T(E,\varphi) = T(E,-\varphi)$ would be restored when the effect from the
neighbourhood of both Dirac cones is included.

The conductivity can then be straightforwardly related to the transmission by
means of the Landauer formula \cite{Landauer:57,Buettiker:86}, as
\begin{equation}
\sigma(E) = \sigma_0 D \frac{E}{\hbar \vF} \int_{-\pi/2}^{\pi/2}
T(E,\varphi) \cos\varphi \, \frac{d\varphi}{2\pi} ,
\label{eq:Landauer}
\end{equation}
where $\sigma_0 = 4 e^2 /h$ is twice the conductance quantum, and the conserved
component of transmitted momentum, \emph{i.e.} that parallel to the barrier, has
been related to the incidence angle through $k_y = E/(\hbar\vF) \sin\varphi$. In
Eq.~(\ref{eq:Landauer}), only the propagating modes have been included in the
integration. One is then prompted to define the adimensional conductivity
\begin{equation}
\sigma_a (E) = \frac{\sigma(E)}{\sigma_0 D\frac{E}{\hbar\vF}} .
\label{eq:sigmaa}
\end{equation}

Fig.~\ref{fig:condsmooth} (right panel) shows the reduced conductivity,
Eq.~(\ref{eq:sigmaa}), as a function of incident energy $E$, for tunneling
across sharp and smooth barriers, Eq.~(\ref{eq:smooth}). One observes
Fabry-P\'erot oscillations, whose amplitude is reduced by increasing smoothing
(\emph{i.e.,} increasing $a$), the barrier then tending to be a more
regular function. Also, the overall increase of the transmission is reflected in
an enhancement of the conductivity. 

In conclusion, we have studied the effect of nonuniform strain on the
conductivity across smooth strain barriers in graphene. While an increase of
smoothing reduces the oscillations of the transmission as a function of the
incidence angle, one finds a reduction of the Fabry-P\'erot oscillations and an
overall enhancement of the conductivity as a function of incidence energy. These
results should help understanding the properties of corrugated graphene.

\bibliographystyle{gHPR}
\bibliography{a,b,c,d,e,f,g,h,i,j,k,l,m,n,o,p,q,r,s,t,u,v,w,x,y,z,zzproceedings,Angilella}
\end{document}